\documentclass[letter]{aa} % for the letters 
\usepackage{graphicx}
\usepackage{txfonts}
\usepackage{natbib}

\begin{document}

   \title{On the correlation between stellar chromospheric flux and the surface gravity of close-in planets}
    \titlerunning{Stellar chromospheric flux and planetary gravity}

%    \subtitle{I. Overviewing the $\kappa$-mechanism}

   \author{A.~F.~Lanza}
   \authorrunning{A.~F.~Lanza}

   \institute{INAF-Osservatorio Astrofisico di Catania, Via S.~Sofia, 78 -- 95123 Catania, Italy\\
             \email{nuccio.lanza@oact.inaf.it}  }

   \date{}

% \abstract{}{}{}{}{} 
% 5 {} token are mandatory
 
  \abstract
   {}
   {The chromospheric emission of stars with close-in transiting planets has been found to correlate with the surface gravity of their planets. Stars with low-gravity planets have on average a lower chromospheric flux.}
   {We propose that this correlation is due to the absorption by circumstellar matter that comes from the evaporation of the planets. Planets with a lower gravity  have a greater mass-loss rate which leads to a higher column density of circumstellar absorption and this in turn explains the lower  level of  chromospheric emission observed in their host stars. We estimated the required column density and found that  planetary evaporation can account for it. We derived a theoretical relationship between the chromospheric emission as measured in the core of the Ca II H\&K lines and the planet gravity. }
   {We applied this relationship to a sample of transiting systems for which both the stellar Ca II H\&K emission and the planetary surface gravity are known and found a good agreement,  given the various sources of uncertainties and the intrinsic variability of the stellar emissions and planetary evaporation rates. We consider  implications for the radial velocity jitter applied to fit the spectroscopic orbits and for the age estimates of planetary systems based on the chromospheric activity level of their host stars. }
   {}


   \maketitle
%
%________________________________________________________________
%
\section{Introduction}
Main-sequence late-type stars have outer convection zones where turbulence excites sound waves. They dissipate in the outermost low-density regions of the atmosphere, leading to the formation of a chromosphere, that is, a layer where temperature increases outwards. In addition to this basal acoustic heating, the energy dissipated by magnetic fields, which are produced by a hydromagnetic dynamo in the convection zone, contributes to the chromospheric emission. In active stars,  magnetic heating dominates  the basal  heating by orders of magnitude, leading to a wide range of chromospheric emission levels in stars with similar effective temperature and mass \citep[e.g., ][]{Ulmschneideretal91}. 

Considering a sample of  transiting planets, \citet{Hartman10} (hereafter H10) and \citet{Figueiraetal14} found an intriguing correlation between the  chromospheric emission of their hosts and the surface gravity of the planets, in which the emission is lower for stars with  planets with a lower surface gravity. 
A suggestion to  explain such a correlation came from  the observations of \citet{Haswelletal12} and \citet{Fossatietal13}. They found a  complete lack of emission in the cores of the  Mg~II~h\&k and Ca~II~H\&K lines in \object{WASP-12} and attributed this to the absorption of circumstellar matter probably coming from  the evaporation of the  planet. 

In this Letter, we propose a mechanism based on the evaporation and subsequent condensation of matter from a close-in planet to explain the correlation found by H10. 
 %__________________________________________________________________
\section{Observations}
We consider transiting exoplanetary systems with both measured chromospheric emission and planetary surface gravity. Chromospheric emission is measured by the $R'_{\rm HK}$ index \citep[e.g.,][]{Knutsonetal10}. It is determined  by computing the chromospheric index $S$, which is the ratio of the  flux measured in the cores of the Ca II H\&K lines using triangular passbands with a FWHM of 1~\AA~to the flux in two reference passbands of 20~\AA~centred on the continuum of the lines. The  $S$ index is converted into a standard scale and  corrected for the residual photospheric emission,  leading to a non-dimensional measure of the chromospheric emission called $R'_{\rm HK}$.  When a given star has several observations, the median value of $\log R'_{\rm HK}$ is adopted. 

The planetary surface gravity $g$ comes directly from the spectroscopic orbit and the transit lightcurve, without  constraints from stellar or atmospheric models \citep{Southworthetal07}. 
%This reduces  systematic errors to a minimum, while the statistical errors are dominated by the uncertainties on the fractional radius of the planet  and the radial-velocity semiamplitude and are of the order of a few percent.  
{%\bf 
We added 15 new transiting planets to the sample of H10 with  gravity and $\log R'_{\rm HK}$ as listed in the exoplanets.org database on 16~October 2014 \citep{Wrightetal11}; this makes a total of 54 systems. The mass $M$ and orbital semimajor axis $a$ of our planets and the effective temperature $T_{\rm eff}$ of their hosts were taken from exoplanets.org. Several planets listed by H10 have no $\log R'_{\rm HK}$ value in that database, therefore  we took $\log R'_{\rm HK}$ from H10 when available, while  $g$ and its uncertainty came from exoplanets.org. For WASP-18, we adopted  $\log R'_{\rm HK} =-5.153$ \citep[see Isaacson, private communication, and][]{Milleretal12}.
 }
\section{Model}
\subsection{Configuration of the stellar coronal field}
In the Sun, prominences form by  condensing  plasma as a results of a thermal instability in long magnetic loops which have a dip close to their top where matter can be accumulated without falling  under the action of gravity. The matter that forms a prominence probably comes from the evaporation of the chromosphere  induced by  magnetic heating that is localized close to the footpoints of the long loop \citep[cf.][and references therein]{Lanzaetal01}. Rapidly rotating late-type stars such as \object{AB~Dor} show signatures of condensations  absorbing in  H$\alpha$, that are located a few stellar radii above the surface \citep{CameronRobinson89a,CameronRobinson89b}. %They are probably  formed by the condensation of plasma in long loops and are stabilized close to their top by the strong centrifugal force that points radially outwards. 
Prominence-like structures were also observed in \object{CoRoT-2} \citep{Czeslaetal12}.

\begin{figure}
   \centering
\includegraphics[width=5.5cm,height=7cm]{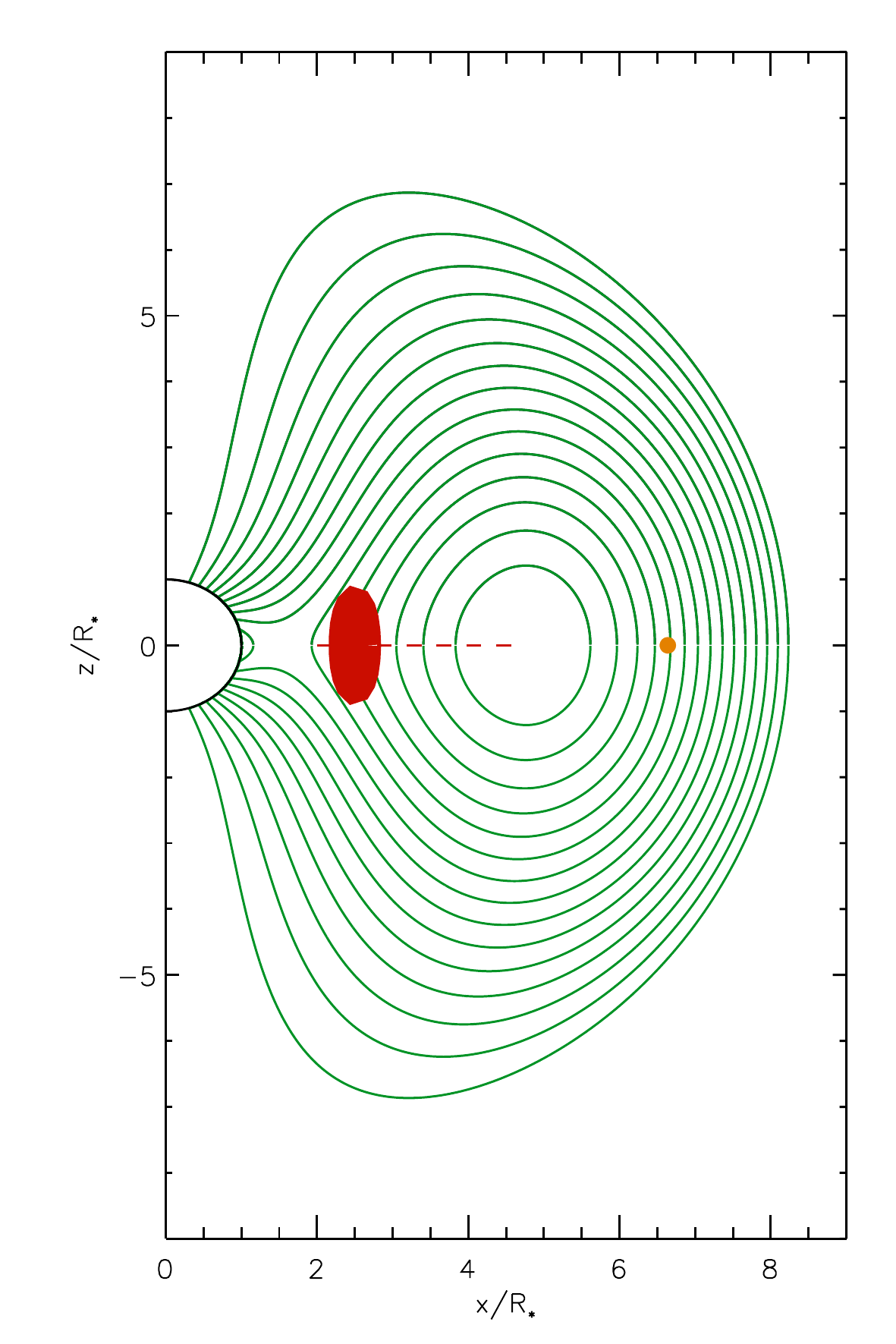}
   \caption{Meridional section of an axisymmetric linear force-free field with an azimuthal flux rope encircling the star.  The  green solid lines are the field lines of the stellar coronal field, the orange dot is a close-in planet from which matter can evaporate and, after moving towards the star along the field lines, condense in the potential well to form a prominence-like structure (in red). The red dashed line on the equatorial plane of the star marks the locus of the minimum gravitational potential along field lines where condensations can be stably stored. The unit of measure is the radius $R_{*}$ of the star, with the $z$ axis being the axis of symmetry of the field and $x$ the distance from the $z$ axis. }
   \label{field_config}
    \end{figure}

In late-type stars that are accompanied by close-in planets, different magnetic configurations are possible, including some  that are capable of sustaining  plasma condensations against the stellar gravitational field. \citet{Lanza09,Lanza12} introduced coronal field models and described their properties. Linear force-free fields  represent the minimum energy configuration for a given total magnetic helicity that is attained thanks to the continuous energy dissipation associated with the reconnection between the coronal and the planetary fields as  the planet orbits the star. Some linear force-free fields have closed field lines that are suitable to store the condensed plasma, as shown in Fig.~\ref{field_config} for an axisymmetric field with some azimuthal twist  \citep[see][for details]{Lanza09}. At a distance of a few stellar radii from the star, there are field lines with dips where the gravitational potential has a relative minimum which allows storing condensed matter. The very long loops make thermal conduction ineffective, so the condensed plasma cannot be re-heated by the hot stellar corona, and the  stellar radiation cannot re-evaporate the matter because the absorption of the Lyman continuum requires column densities orders of magnitude greater than expected in those structures (see Sect.~\ref{column_density}). Therefore, it is conceivable that, once formed, plasma condensations at chromospheric temperatures can remain stable. The weight of the stored material helps to stabilize the whole configuration because a sizable change of the field line geometry would lift up the matter which in turn would  require additional energy \citep[cf.][]{Lanza09}. For the same reason, large quiescent solar prominences are the most long-lived structures in the solar corona with lifetimes that can reach ten months. 
%%%%%%%%%%%%%%%%%%%%%%%%%%%%%%%%%%%%%%%%%%%%%%%%%%%%%%%
    
%%%%%%%%%%%%%%%%%%%%%%%%%%%%%%%%%%%%%%%%%%%%%%%%%%%
%
\subsection{Absorption of the chromospheric flux by circumstellar condensations}
\label{column_density}

As observed in the Sun,  the thermal instability by which  condensations are formed produces clumps of relatively low-temperature material that is separated by regions of higher temperature and lower density in approximately  pressure equilibrium. Therefore, we do not expect that the azimuthal flux rope encircling the star is uniformly filled with a homogeneous condensation, but that several clumps of material are formed inside it. They absorb at different wavelengths because of their individual radial velocities. Since the matter comes from the evaporating planet, it initially has its orbital angular momentum. To allow it to condense closer to the star inside a coronal field that rotates with the stellar angular velocity, this matter must give up some of its initial angular momentum which is transferred to the star by  magnetic stresses. However, since the rope magnetic field  has an azimuthal component, the condensed plasma can flow in the azimuthal direction  along the magnetic field lines and acquire azimuthal velocities up to several tens of km~s$^{-1}$ thanks to the  initial angular momentum it had when it left the planet. In conclusion, we expect to find prominences capable of absorbing all along the 1~\AA~ intervals centred on the cores of the Ca II H\&K lines used to measure the chromospheric index. 

\citet{Vial82a, Vial82b} observed and modelled several intense lines in solar prominences. We adopted his radiation transfer model to estimate the optical depth $\tau_{0}$  at the central wavelengths of the Ca II H\&K lines across a slab of condensed matter of thickness $X$, with a typical temperature of $8 \times 10^{4}$~K, electron density $2 \times 10^{10}$ cm$^{-3}$, and a hydrogen ionization fraction $n_{\rm p}/ n_{1} = 3$, where $n_{\rm p}$ is the number density of the protons and $n_{1}$ that of the hydrogen atoms in the fundamental state. A microturbulence velocity of 8~km~s$^{-1}$ was adopted.  Assuming solar abundance for Ca, the hydrogen column density corresponding to $\tau_{0} = 1$ in the core of the Ca II K line is $N_{\rm K} = 1.8 \times 10^{18}$~cm$^{-2}$, while for the Ca II H line it is $N_{\rm H} = 3.5 \times 10^{18}$ cm$^{-2}$. The FWHM of the lines is $\sim 0.2$~\AA,~implying that at least five absorbing clumps with radial velocities within $\pm 40$ km~s$^{-1}$ of the stellar radial velocity are needed to uniformly absorb along the 1~\AA~ windows where the chromospheric emission is measured. 

The observed range of $R'_{\rm HK}$ in our sample of stars covers about one order of magnitude (cf. Sect.~\ref{results} and Fig.~\ref{rhkvsg}) implying that the highest absorption of the flux is about a factor of ten, corresponding to an average optical depth across the 1~\AA~ windows of $\tau \simeq 2.5$. This implies a total hydrogen column density, considering all the five different clumps absorbing at different radial velocities, of $\sim 4.4 \times 10^{19}$~cm$^{-2}$ for the Ca II H line and half that value for the Ca~II~K line. We estimated the total amount of matter into the condensations by assuming that it is stored in a cylindrical slab of base radius $r_{\rm T}$ that extends in the direction parallel to the axis of symmetry for at least $2R_{*}$, where $R_{*}$ is the radius of the star, to completely cover the stellar disc. Therefore, the volume filled with the condensations is approximately $V = 4 \pi\, r_{\rm T}\, R_{*}\, X$, where $X \ll R_{*}$ is their typical thickness. Following Fig.~\ref{field_config}, we assume $r_{\rm T} \sim 2.5\, R_{*}$. For $X = 5 \times 10^{8}$~cm, we need an average hydrogen density of $n = 8.8 \times 10^{10}$~cm$^{-3}$ to account for the highest total absorption. For an F-type star, $R_{*} \sim 10^{11}$~cm, hence $V \sim 1.6 \times 10^{32}$~cm$^{3}$ and the total highest condensation mass $M_{\rm c} \sim 2.3 \times 10^{19}$~g.  If $M_{\rm c}$ is provided by the evaporation of the planet during, for example, ten months, the required evaporation rate is $\dot{M} \sim 8.8 \times 10^{11}$~g~s$^{-1}$, which is at the upper limit of the estimated evaporation rates for close-in planets \citep[cf. ][]{Lecavelierdesetangs07}. Since stars hosting hot Jupiters and the atmospheres of their planets are generally richer in metals than the Sun \citep[e.g., ][]{Fortneyetal06}, we can reduce the required hydrogen column density and the planetary evaporation rate by a factor of $\sim 2-10$. For planets closer than $a \la 0.05$~AU, the energy released by magnetic reconnection between the stellar and the planetary fields can increase the evaporation rate up to $10^{12}-10^{13}$~g~s$^{-1}$, providing the required amount of matter on a timescale of the order of a few months \citep[cf.][]{Lanza13}. 
\subsection{Dependence of the stellar chromospheric flux}
\label{chromospheric_flux}
Assuming an absorption with optical depth $\tau$ uniform across the 1~\AA~ windows used to measure the chromospheric flux, the observed chromospheric index is given by 
\begin{equation}
\log R'_{\rm HK} = \log R'^{(0)}_{\rm HK} - 0.434\, \tau,
\label{eq1}
\end{equation}
where $R'^{(0)}_{\rm HK}$ is the index one should measure in the absence of absorption. The line optical depth is given by  
\begin{equation}
\tau = \alpha n X, 
\label{eq2}
\end{equation}
where $\alpha$ is the  absorption coefficient per hydrogen atom (in cm$^{2}$), $n$ the hydrogen number density (in cm$^{-3}$), and $X$ the geometric depth (in cm). The optical depth in the continuum reference passbands is  $\la 2 \times 10^{-6}$ of that in the line cores \citep[cf. Table 2 in][]{Vial82b} and is thus completely negligible. 

The density $n$ is given by 
\begin{equation}
n = \frac{\dot{M} \, t_{\rm a}}{m_{\rm p} V},
\label{eq3}
\end{equation}
where $t_{\rm a}$ is the time for which the evaporated matter is accumulated, {%\bf 
$V = 4\pi \, r_{\rm T}\, R_{*}\, X$ the volume of the condensation region as specified in Sect.~\ref{column_density},} and $m_{\rm p} = 1.67 \times 10^{-24}$~g  the proton mass. The accumulation time $t_{\rm a}$ is comparable to the time that the evaporating matter takes to travel from the planet to the condensation site. 
Therefore, assuming that matter diffuses along the coronal field lines under the action of turbulent diffusion with a coefficient $\nu$, we have 
\begin{equation}
 t_{\rm a} \sim a^{2}/\nu \sim a^{2}/(c_{\rm s} R),
 \label{eq4} 
\end{equation}
 where $a$ is the distance of the planet from the star on a circular orbit and  $\nu \sim c_{\rm s} R$, where $c_{\rm s}$ is the sound speed and $R$ the radius of the planet. This expression comes from a  dimensional argument by considering that the evaporation speed  is of the order of $c_{\rm s}$ \citep[e.g.,][]{Adams11} and the turbulence is induced by the orbital motion of the planet with a typical lengthscale  $\sim R$. {%\bf 
This scaling for $t_{\rm a}$ makes {the density $n$ in  Eq.~(\ref{eq3})} independent of  $a$ (see below), however, other scalings are plausible (cf. Appendix~\ref{appendixA}). We favour the expression in Eq.~(\ref{eq4}) because the statistical  correlation between $\log R'_{\rm HK}$ and $a$ has a low significance (cf. Sect.~\ref{results}). }

The mass-loss rate is powered by the stellar extreme-ultraviolet flux $F_{\rm EUV}$ \citep[cf. ][]{Lecavelierdesetangs07} and can be estimated by assuming that the flux received by the planet $\pi F_{\rm EUV} (R_{*}/a)^{2} R^{2}$ is converted into work against its gravitational field with an efficiency $\eta$ 
{ \citep[e.g., ][]{Watsonetal81,Sanz-Forcadaetal11}}, that is, 
\begin{eqnarray}
\dot{M} & = & \pi \, \eta \, F_{\rm EUV} \left( \frac{R_{*}}{a} \right)^{2} \frac{R^{3}}{GM} \nonumber \\
            & = & \pi \, \eta \,  F_{\rm EUV} \left( \frac{R_{*}}{a} \right)^{2} g^{-1} R, 
\label{eq5}            
\end{eqnarray}
where $G$ is the gravitation constant and we have explicitly introduced the surface gravity  $g = GM/R^{2}$. Substituing Eqs.~(\ref{eq5}), (\ref{eq4}), and (\ref{eq3}) into Eq.~(\ref{eq2}), we obtain an expression for the mean optical depth $\tau$ and recast Eq.~(\ref{eq1}) as 
\begin{equation}
\log R'_{\rm HK} = \log R'^{(0)}_{\rm HK} - \gamma g^{-1},
\label{relationrhkg}
\end{equation}
where  $\gamma \equiv 0.0434\, (\alpha\, \eta\, F_{\rm EUV})/(m_{\rm p} c_{\rm s})$ depends on the stellar high-energy radiation and the temperature of the evaporation flow that determines the sound speed. 
 \section{Results}
 \label{results}
\begin{figure}
\centerline{
\includegraphics[width=6cm,height=9cm,angle=90]{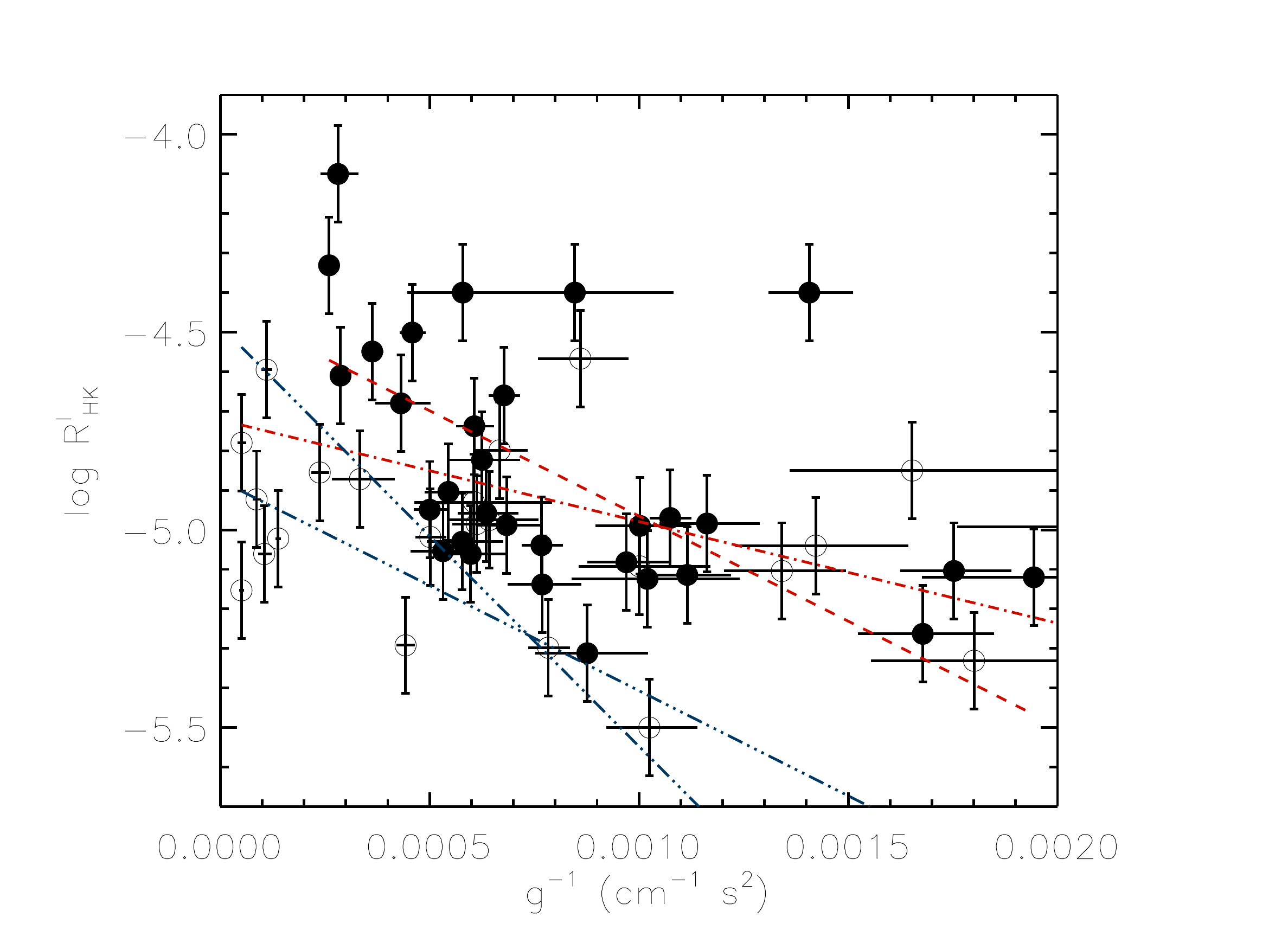}}
   \caption{Chromospheric emission index $R'_{\rm HK} $ vs. the inverse of the surface planetary gravity $g$  for our sample of close-in planets.  The red dashed line is a linear regression computed by considering the 31 data points (filled circles) with $M > 0.1$~M$_{\rm J}$, $a < 0.1$~AU and stellar effective temperature $4200 < T_{\rm eff} < 6200 $~K. The red dot-dashed line is a linear regression computed  with the entire set of 54 datapoints. Those not included in the restricted sample are indicated by open circles. Note that the point corresponding to WASP-17b (an open circle) is outside the scale of the plot, but it is included in the regression. The three-dots-dashed blue lines are the theoretical relationship (\ref{relationrhkg}) for the Sun (the line with the lower slope) and for the illustrative case of a star with twice the solar chromospheric and $F_{\rm EUV}$ fluxes.  }
   \label{rhkvsg}
    \end{figure}

In Fig.~\ref{rhkvsg}, we plot the stellar chromospheric index $\log R'_{\rm HK}$ vs. the inverse of the surface gravity of the planet. 
{%\bf 
H10 considered a subsample with $M> 0.1$~M$_{\rm J}$, $a < 0.1$~AU and $4200 < T_{\rm eff} < 6200$~K because they have calibrated values of $R'_{\rm HK}$ and better determined parameters (his subsample 1).  We used the same criteria for selecting our corresponding subsample, finding 31 planets that are plotted in Fig.~\ref{rhkvsg} with filled dots. The red dashed line is the linear regression 
\begin{equation}
\log R'_{\rm HK} = -(4.43 \pm 0.07) - (532 \pm 78)\, g^{-1},
\label{eq7}
\end{equation}
where $g$ is in cm~s$^{-2}$ and the uncertainties  correspond to one standard deviation. They were computed by assuming a standard deviation of 0.12 for $\log R'_{\rm HK}$ as a result of measurement and conversion errors \citep{Knutsonetal10}, and stellar activity cycles.  In the Sun, the peak-to-peak variation along the 11-yr cycle amounts to $\Delta \log R'_{\rm HK} \sim 0.24$ \citep{Baliunasetal95} and  can be assumed as typical for stars with a low activity level which is the case of most of our planet hosts. 

The Spearman correlation coefficient is $-0.61$ with a  probability of chance occurrence of 0.030 percent. We obtain a chi square of 124.4 for the best fit in Eq.~(\ref{eq7}). The fit residuals are higher than expected on the basis of the data uncertainties and can be explained by considering the differences in the intrinsic EUV flux, and, to a lesser extent, in the evaporation flow temperatures of different planets that make the parameter $\gamma$ remarkably variable from one planet to the other. For example, in the Sun, $F_{\rm EUV}$ varies by a factor of $\approx 3-4$ along the 11-yr cycle, and  larger variations are observed during major stellar flares, which affect planetary evaporation rates \citep{Lecavelierdesetangsetal12}.  
 We estimate the mean $F_{\rm EUV}$ from the  slope $\gamma = 532$ cm~s$^{-2}$, adopting $\alpha = 2.9 \times 10^{-19}$ cm$^{-2}$, corresponding to solar abundance, $\eta = 0.5$, and $c_{\rm s} = 15$ km~s$^{-1}$, that is, a temperature of $\sim 10^{4}$~K for the evaporation flow. We find $F_{\rm EUV} =  4.7 $ erg cm$^{-2}$ s$^{-1}$ at  the distance of 1~AU, similar to the mean $F_{\rm EUV}$ of the Sun \citep[cf. ][]{Lecavelierdesetangs07}. Note that our model implicitly assumes $R'^{(0)}_{\rm HK}$ to be the same for all the stars, so the correlation should become stronger if $R'^{(0)}_{\rm HK}$ were individually known together with $F_{\rm EUV}$ and $c_{\rm s}$. 

We also computed a linear regression for our complete sample of 54 data points, corresponding to subsample 3 of H10. It is plotted as a dot-dashed line in Fig.~\ref{rhkvsg} with intercept $-4.72 \pm 0.04$ and slope $\gamma = - 257 \pm 43$ cm~s$^{-2}$.  The remarkable variation of the regression line parameters arises because  several datapoints fall below the regression~(\ref{eq7}) for $g^{-1} < 0.0005$~cm$^{-1}$~s$^{2}$.

To allow for a preliminary comparison with our theory, we plot  Eq.~(\ref{relationrhkg}) in Fig.~\ref{rhkvsg} with the above constant values for $\alpha$, $\eta$, $c_{\rm s}$, and assuming mean solar flux values, that is  $\log R'^{(0)}_{\rm HK} = -4.875$ and $F_{\rm EUV} = 4.64$ erg cm$^{-2}$ s$^{-1}$ at 1~AU, or twice solar flux values, to illustrate the case of a sun-like or of a more active star, respectively. The intercepts of those theoretical regression lines are smaller than those found by fitting the data points, suggesting that most of the sample stars  are remarkably more chromospherically active than the Sun. On the other hand, the slopes are similar, that is, their average $F_{\rm EUV}$ is similar to the Sun's.

H10 and \citet{Figueiraetal14} plotted $\log R'_{\rm HK}$ vs. $\log g$ instead of our physically motivated relationship. The chi square of the linear regression between $\log R'_{\rm HK}$ and $\log g$ is 117.1 for subsample 1, giving no significant preference for this functional form with respect to that we adopted. 

Finally, the Spearman correlation coefficient between $\log R'_{\rm HK}$ and $a$ is $-0.19$ corresponding to a probability of a chance association of $\sim 30$ percent (cf. Sect.~\ref{chromospheric_flux} and Appendix~\ref{appendixA}). }

%%%%%%%%%%%%%%%%%%%%%%%%%%%%%%%%%%%%%%%%%%%%%%%%%
%%%%%%%%%%%%%%%%%%%%%%%%%%%%%%%%%%%%%%%%%%%%%%%%%%   
 %
 \section{Discussion and conclusions}
 
 We propose an explanation for the correlation between the stellar chromospheric emission and the surface gravity of close-in exoplanets  \citep[H10 and][]{Figueiraetal14}. It is based on  the absorption by prominence-like structures around the host star formed by the plasma evaporating from the planet \citep{Lanza09}. The observations of WASP-12 by \citet{Haswelletal12} and \citet{Fossatietal13} provide support to the proposed mechanism. WASP-12 is an extreme system in terms of closeness of its hot Jupiter to its Roche lobe, which should strongly enhance its evaporation rate. This  agrees with the fact that its $\log R'_{\rm HK}$ is the lowest of our sample.  
 
{%\bf 
The possibility that a low  $R'_{\rm HK}$  indicates an intrinsic low activity level in some of our stars cannot be excluded.  \citet{Pillitterietal14} found that circumstellar absorption cannot explain the low level of  X-ray emission in \object{WASP-18}, contrary to the suggestion by \citet{Fossatietal14}. The strong tides generated on the star by such a massive and close-in planet might significantly affect the stellar dynamo and remarkably reduce its activity. {On the other hand, evidence for hot Jupiters spinning up their hosts} was found by \citet{PoppenhaegerWolk14} in CoRoT-2 and possibly \object{HD~189733}. 

The evolution of the intrinsic chromospheric emission  is significant only during the first $\sim 2$~Gyr on the main sequence \citep{Pace13}, which means that it does not greatly affect our sample that mainly consists of solar-age stars.  Moreover, planetary mass loss through evaporation, even at a rate of $10^{12}$ ~g~s$^{-1}$, has a limited effect on Jupiter-mass objects. We refer to H10 for more considerations on evolutionary effects.}

A consequence of our  model  is that a measure of the chromospheric emission cannot be immediately taken as a proxy for the photospheric activity level of a planet-hosting star. The intrinsic activity can be significantly higher than derived from $R'_{\rm HK}$, and this can affect the estimate of the radial-velocity jitter term  included when fitting the spectroscopic orbit of a transiting planet. Age estimates based on the chromospheric emission level \citep[e.g.,][]{MamajekHillebrand08} can lead to systematically older values than the true ages for stars with low-gravity planets. 

We implicitly assumed alignment between the stellar spin, the axis of symmetry of the coronal field that supports the condensations, and the orbital angular momentum. 
However, a misalignment between the stellar spin and the orbital angular momentum is observed in several systems, particularly when $T_{\rm eff} \ga 6200$~K \citep{Albrechtetal12}. This suggests that  systems with an inclination  different from $90^{\circ}$ may also display effects of circumstellar absorption. 
\begin{acknowledgements}
The author wishes to thank the anonymous referee for several valuable comments that helped him to improve this work. He is grateful to S.~Desidera, C.~Haswell, H.~Isaacson, I.~Pillitteri, K.~G.~Strassmeier, and S.~Wolk for interesting discussions and acknowledges support by INAF through the {\it Progetti Premiali} funding scheme of the Italian Ministry of Education, University, and Research. 
\end{acknowledgements}

\appendix
\section{Alternative scalings for the optical depth}
\label{appendixA}
{%\bf 
Alternative scalings to those adopted in Sect.~\ref{chromospheric_flux} can be considered. In our hypotheses, the total volume $V$ is independent of the orbit semimajor axis $a$ because the size of the region enclosing the relative minimum of the gravitational potential along the magnetic field lines scales with the radius of the star $R_{*}$, given that the boundary conditions controlling the field geometry are fixed at the surface of the star. Therefore, we expect that in general the vertical extension of the slab depends on $R_{*}$ rather than the distance $a$ of the planet. In other words, the volume $V$ should be proportional to $R_{*}^{2}X$ although the coefficient of proportionality is not necessarily that adopted above, that is, $4\pi\, \times 2.5$, which corresponds to the smallest vertical extension of the slab $\sim 2 R_{*}$. 

The orbit semimajor axis $a$ controls the evaporation rate $\dot{M}$ and the timescale $t_{\rm a}$ for the filling of the potential well along the field lines. With the scaling adopted above, these two dependences cancel  each other in the final expression for the mean density $n$, but other scalings for $t_{\rm a}$ are plausible and  generally lead to a dependence of $n$ on $a$. 

The shortest possible timescale corresponds to the free-fall time in the case of negligible pressure gradients and Lorentz force, and neglecting the ram pressure of the stellar wind. { In this case, $t_{\rm a} = \sqrt{2} \pi (GM_{*})^{-1/2} a^{3/2}$}, where $M_{*}$ is the mass of the star, and the optical depth appearing in Eq.~(\ref{eq1}) becomes 
\begin{equation}
\tau = \sqrt{2} \pi \frac{\alpha \,  \eta \, F_{\rm EUV}}{10\, m_{\rm p}} \sqrt{\frac{M}{M_{*}}} a^{-1/2} g^{-3/2}.
\end{equation}
Conversely, the longest timescale is that given by the diffusion process considered in Sect.~\ref{chromospheric_flux}. An intermediate possibility is that of a syphon flow along the magnetic field lines that directly carries the evaporated matter into the potential well. The typical velocity of that flow is of the order of the sound speed $c_{\rm s}$ \citep[cf. the evaporation model in the presence of planetary and stellar magnetic fields by][]{Adams11}, thus $t_{\rm a} \sim a / c_{\rm s}$. With that scaling, we find 
\begin{equation}
\tau \sim \frac{\alpha \, \eta \, F_{\rm EUV }}{10 \, m_{\rm p} \, c_{\rm s}} \left( \frac{R}{a} \right) g^{-1},  
\end{equation}
where $c_{\rm s}$ is constant in the case of a steady isothermal flow. 

}
\end{document}